\documentclass[Danish,UKEnglish]{llncs}
\usepackage[applemac]{inputenc}

\usepackage{stackengine}
\newcommand\ttleq{\raisebox{1pt}{$\mathrel{\stackunder[-1.5pt]{\texttt{<}}{\texttt{-}}}$}}

\usepackage{enumerate}
\usepackage{paralist}
    \usepackage{lineno,hyperref}
\usepackage{amsmath}
\usepackage{amssymb}
\usepackage{graphicx}
\usepackage{url}
\usepackage[all]{foreign}

\usepackage{todonotes}
\usepackage[caption=false]{subfig}
\usepackage{tikz}
\usetikzlibrary{positioning,shapes,fit,arrows.meta,decorations}
\usetikzlibrary{shapes.geometric}
\usetikzlibrary{decorations.pathreplacing,decorations.pathmorphing}

\def\rightarrowmetalogic{\mathrel{\ourmapsto\supmeta\sublogic}}

\def\true{\mathit{true}}

\def\failure{\mathit{failure}}

\def\where{\mathrel{\mbox{\textsc{where}}}}

\def\mapsfrom{\mathrel{\mbox{$\leftarrow$\hskip -0.2ex\vrule width 0.15ex height 1ex\hbox to 0.3ex{}}}}

\def\capplus{\mathop{\setbox1=\hbox{$\cap$}\vbox to \ht1{\hbox{$\cap$}\vss\hbox to \wd1{\hfil\scriptsize$+$\hfil}}}}
\def\ourmapsto{\mathrel{\mbox{\hbox to 0.2ex{}\vrule width 0.15ex height 1ex\hskip -0.2ex$\rightarrow$}}}
\def\ourmapsfrom{\mathrel{\mbox{\hbox to 0.2ex{}$\leftarrow$\hskip -0.2ex}\vrule width 0.15ex height 1ex}}
\def\ourlongmapsto{\mathrel{\mbox{\hbox to 0.2ex{}\vrule width 0.15ex height 1ex\hskip -0.2ex$\longrightarrow$}}}

\def\mapstostar{\stackrel{*}\ourmapsto}
\def\mapsfromstar{\stackrel{*}\mapsfrom}

\def\supmeta{^{\mathit{meta}}}

\def\sublogic{_{\mathit{logic}}}

\def\approxsub#1{\mathrel{\approx_{#1}}}
\def\approxlogic{\mathrel{\approx\sublogic}}
\def\approxmetalogic{\mathrel{\approx\sublogic\supmeta}}
\def\approxsup#1{\mathrel{\approx^{#1}}}

\def\Mmetalogic{\mathcal{M}}
\def\Imetalogic{I\supmeta\sublogic}
\def\Ilogic{I\sublogic}
\def\Imetalogic{I\supmeta\sublogic}

\def\cinv{\text{{\rm{\texttt{inv}}}}}
\def\cequiv{\text{{\rm{\texttt{equiv}}}}}

\def\denotes#1{[\![#1]\!]}
\def\denotesgr#1{[\![#1]\!]^{\mathit{Gr}}}

\def\denoteslogiclogic#1{[\![#1]\!]_{\mathit{logic}}^{\mathit{logic}}}

\def\denotesmetalogic#1{[\![#1]\!]_{\mathit{logic}}^{\mathit{meta}}}

\def\rightarrowsub#1{\mathrel{\ourmapsto_{#1}}}
\def\rightarrowsublogic{\mathrel{\ourmapsto_{\mathit{logic}}}}
\def\leftarrowsublogic{\mathrel{\ourmapsfrom_{\mathit{logic}}}}

\def\rightarrowsup#1{\mathrel{\ourmapsto^{#1}}}

\def\ourlongmapstometalogic{\mathrel{\mbox{\hbox to 0.2ex{}\vrule width 0.15ex height 1ex\hskip -0.2ex$\longrightarrow$}\sublogic\supmeta}}
\def\ourlongmapsfrommetalogic{\mathrel{\mbox{\hbox to 0.2ex{}$\longleftarrow$\hskip -0.3ex\vrule width 0.15ex height 1ex}\sublogic\supmeta}}

\def\eyes{\hbox to 1.84ex{\vbox to 0pt{\vss\hbox to 1.84ex{\hfil\vbox to 0pt{\hbox{\textrm{.}}}\hfil\vbox to 0pt{\hbox{\textrm{.}}\vss}\hfil}}}}
\def\ssmile{\hbox to 1.84ex{\vbox to 0pt{\vss\hbox to 1.84ex{\hfil\vbox to 0pt{\vss\hbox{\tiny$\smile$}\vss}\hfil}}}}
\def\ffrown{\hbox to 1.84ex{\vbox to 0pt{\vss\hbox to 1.84ex{\hfil\vbox to 0pt{\vss\hbox{\tiny$\frown$}\vss}\hfil}}}}
\def\bvadr{\hbox to 1.84ex{\vbox to 0pt{\vss\hbox to 1.84ex{\hfil\vbox to 0pt{\vss\hbox{\tiny$\leftrightsquigarrow$}\vss}\hfil}}}}

\def\willsucced{\hbox{\hskip 0.161ex\vrule height 1.84ex\vbox to 1.84ex
        {\hrule width 1.84ex\vskip 0.391ex\eyes\vskip -1.886ex\ssmile\vss\hrule width 1.84ex}\vrule height 1.84ex}\hskip 0.18ex}
\def\willfail{\hbox{\hskip 0.161ex\vrule height 1.84ex\vbox to 1.84ex
        {\hrule width 1.84ex\vskip 0.391ex\eyes\vskip -1.886ex\ffrown\vss\hrule width 1.84ex}\vrule height 1.84ex}\hskip 0.18ex}

\def\namedrule#1#2#3#4#5{#1\colon #2 \text{\rm\texttt{\char92}}#3\, \text{\rm\texttt{<=>}} \,#4 \text{\rm\texttt{|}} #5}
\def\nonamerule#1#2#3#4{#1 \text{\rm\texttt{\char92}}#2\, \text{\rm\texttt{<=>}} \,#3 \text{\rm\texttt{|}} #4}

\title{Confluence of CHR revisited:  
invariants and modulo equivalence\thanks{This work is supported by The Danish Council for Independent Research, Natural Sciences,      grant no.~DFF 4181-00442.}
}
\titlerunning{Confluence of CHR revisited} 

\author{Henning Christiansen \and Maja H.~Kirkeby}
\institute{Computer Science, Roskilde University, Denmark\\ \email{henning@ruc.dk} and \email{majaht@ruc.dk}}

\begin{document}
\maketitle
\begin{abstract}
Abstract simulation 
of one transition system by another
is introduced as a means to simulate a potentially infinite class of similar transition sequences within a single transition sequence.
This is useful for proving confluence under invariants of a given system, as it 
may reduce the number of proof cases to consider  from infinity to a finite number.
The classical confluence results for Constraint Handling Rules (CHR)
can be explained in this way, using CHR as a simulation of itself.
Using an abstract simulation based on a ground representation,
we extend these results to include
confluence under invariant
and modulo equivalence, which have not been done in a satisfactory way before.
\end{abstract}



\section{Introduction}
Confluence of a transition system means that any two alternative transition sequences from a given state can be
extended to reach a common state.
Proving confluence of nondeterministic systems may be important for correctness proofs and it  anticipates  parallel implementations
and application order optimizations.
%
Confluence modulo equivalence generalizes this so that these ``common states'' need not be identical, but only equivalent according to an equivalence relation. 
This allows for redundant data representations (e.g., sets as lists) and procedures that search for an optimal
solution to a problem, when any of two equally good solutions can be accepted (e.g., the Viterbi algorithm
analyzed for confluence modulo equivalence in~\cite{DBLP:journals/fac/ChristiansenK17}).


We introduce a notion of abstract simulation of one system, the object system, by another, the meta level system,
and show how proofs of confluence (under invariant, modulo equivalence) for an object system may be 
expressed within a meta level system. 
This may reduce the number of proof cases to be considered, often from infinity to a finite number.
We apply this to the programming language of Constraint Handling Rules, CHR~\cite{DBLP:conf/iclp/Fruhwirth93,fruehwirth-98,fru-chr-book-2009},
giving a clearer exposition of existing results
and extending them
for
invariants and modulo equivalence.
%

By nature, invariants and state equivalences are meta level properties that in general cannot be expressed
in its own system: the state itself is implicit and properties such as groundness
(or certain arguments restricted to be uninstantiated variables)
cannot be expressed in a logic-based semantics for CHR.
Using abstract simulation we can add the necessary enhanced expressibility to the meta level,
and the ground representation of logic programs, that was studied in-depth in the late 1980s and -90s in the context of meta-programming
in logic (e.g.,~\cite{DBLP:journals/jlp/Christiansen98,DBLP:conf/meta/HillL88,Hill94meta-programming-in-logic}), comes in readily as a well-suited and natural choice for this.
%
%
%
The following minimalist example motivates both invariant and state equivalence for CHR.
\begin{example}
[\cite{DBLP:conf/lopstr/ChristiansenK14,DBLP:journals/fac/ChristiansenK17}]\label{ex:set-informal}
The following CHR program, consisting of a single rule, collects a number of separate items
into a set represented as a list of items.
\begin{verbatim}
set(L), item(A) <=> set([A|L]).
\end{verbatim}
This rule will apply repeatedly, replacing constraints matched by the left hand side by the one indicated to the right.
The query
\begin{verbatim}
?- item(a), item(b), set([]).
\end{verbatim}
may lead to two  different final states, $\{\texttt{set([a,b])}\}$ and  $\{\texttt{set([b,a])}\}$,
both representing the same set.
Thus, the program is not confluent, but it may be
confluent modulo an equivalence that
disregards the order of the list-elements.
Confluence modulo equivalence still requires an invariant that excludes more than one \texttt{set}/1 constraint, as otherwise, 
an element may go to an arbitrary of those.
\end{example}

\subsection{Related work}
Some applications of our abstract simulations may be seen as special cases
of abstract interpretation~\cite{DBLP:conf/popl/CousotC77}.
This goes for the re-formulation of
the classical confluence results for CHR, but when invariants are introduced, this is not obvious;
a detailed argument is given in  Section~\ref{sec:conflLogicBased}.
%
It is related to symbolic execution and constraint logic
programming~\cite{DBLP:conf/popl/JaffarL87}, where reasoning takes
place on compact abstract representations parameterized in suitable ways, rather than checking multitudes of concrete instances.
Bisimulation~\cite{DBLP:conf/tcs/Park81}, which has been applied in many contexts,  indicates a tighter relationship between states and transitions of two systems than the abstract simulation: when a state $s_0$ is simulated by an abstract state $s_0'$ and there is a transition $s_0 \rightarrow s_1$, bisimulation would require the existence of an abstract transition $s_0' \rightarrow' s_1'$, which may not be case  as demonstrated by Example~\ref{ex:zigzagPlainCont}. 
%

Previous results on confluence of CHR programs, e.g.,~\cite{DBLP:conf/cp/Abdennadher97,DBLP:conf/cp/AbdennadherFM96,DBLP:journals/constraints/AbdennadherFM99}, mainly refer to a logic-based semantics,
which is well-suited for showing program properties, but it does not comply with typical implementations~\cite{DBLP:journals/aai/HolzbaurF00a,SchrijversDemoen2004}
and applies only for a small subset of CHR programs.
Other works~\cite{DBLP:conf/lopstr/ChristiansenK14,DBLP:journals/fac/ChristiansenK17} 
suggest an alternative operational semantics that lifts these limitations, including the ability to handle
Prolog-style built-in predicates such as \texttt{var}/1,   etc. 
To compare with earlier work and for simplicity, the present paper
refers to the logic-based semantics.

As long as invariants and modulo equivalence are not considered, 
 the logic-based semantics allows for elegant confluence proofs based on Newman's Lemma (Lemma~\ref{lm:newman}, below).
A finite set of critical pairs can be defined, whose joinability ensures confluence for terminating programs.
Duck et al.~\cite{DBLP:conf/iclp/DuckSS07}
proposed a generalization of this approach 
to confluence under invariant, called observable confluence;
no practically relevant methods were suggested, and (as the authors point out) even a simple
invariant such as groundness explodes into infinitely many cases. 

Confluence modulo equivalence was
introduced and motivated for CHR by~\cite{DBLP:conf/lopstr/ChristiansenK14},
also arguing that invariants are important for specifying meaningful equivalences.
An in-depth theoretical analysis, including the use of a ground representation,
is given by~\cite{DBLP:journals/fac/ChristiansenK17}  in relation the alternative semantics mentioned above.
However, it has not been related to abstract simulations,  and the proposal for a
detailed language of meta level constraints in the present paper is new.
Repeating the motivations of~\cite{DBLP:conf/lopstr/ChristiansenK14,DBLP:journals/fac/ChristiansenK17}
in the context of the logic-based semantics,
\cite{GallFr2018} suggested to handle confluence modulo equivalence
along the lines of~\cite{DBLP:conf/iclp/DuckSS07}, thus inheriting the problems  of
infinitely many proof case pointed out above.

An approach to show confluence of a transition system, by producing a mapping into
another confluent system, is described by~\cite{CurienGhelli1991} and
extended to confluence modulo equivalence by~\cite{KirkebyChristiansenK18}; the relationship between such two systems is
different from the abstract simulations introduced in the present paper.
Confluence, including modulo equivalence, has been studied since the first half of the 20th century in a variety of contexts; 
see, e.g.,~\cite{DBLP:journals/fac/ChristiansenK17,DBLP:journals/jacm/Huet80} for overview.

\subsection{Contributions}
We introduce abstract simulation as a setting for proofs of confluence for general
transitions systems and demonstrate this specifically
for CHR.
We recast classical results (without invariant and equivalence), showing
that they are essentially based on a simulation of CHR's logic-based semantics by itself,
and we can pinpoint, why it does not generalize for invariants (see Example~\ref{ex:zigzagPlain}, p.~\pageref{ex:zigzagPlain}).

These results are extended for invariants and modulo equivalence,
using an abstract simulation; it is based on a ground meta level representation
and suitable meta level constraints to reason about it.

\subsection{Overview}
Sections~\ref{sec:basics} and~\ref{sec:simulation} introduce basic concepts of confluence
plus our notion of abstract simulation. 
Section~\ref{sec:chr} gives syntax and semantics of CHR along with a discussion of how much nondeterminism to include
in a semantics used when considering confluence.
Section~\ref{sec:conflLogicBased} re-explains
the classical results in terms of abstract simulation. 
Section~\ref{sec:inf-mod-eq} extends these results for invariants and modulo equivalence;
proofs can be found in an extended report~\cite{ConfModEqforCHRRevisited}.
The concluding Section~\ref{sec:conclusion} gives a summary and explains briefly
how standard mechanisms, used to prevent loops by CHR's propagation rules, can be added.


\section{Basic concepts, confluence, invariants and equivalences}\label{sec:basics}
A \emph{transition system} $D=\langle S,\ourmapsto\rangle$ consists of
a set of \emph{states} $S$, and a \emph{transition} is an element of $\ourmapsto\colon S\times S$,
written  $s_0\ourmapsto s_1$ or, alternatively, $s_1\mapsfrom s_0$.
A \emph{transition sequence} or \emph{path} is a chain of transitions $s_0\ourmapsto s_1\ourmapsto\cdots\ourmapsto s_n$
where $n\geq 0$; if such a path exists, we write $s_0\mapstostar s_n$.
A state $s_0$ is \emph{final} (or \emph{normal form}) whenever \hbox{$\nexists s_1\; s_0\ourmapsto s_1$}, and
$D$ is \emph{terminating} whenever every path is finite.
To anticipate the application for logic programming systems, a given transition system may
have a special final state called $\failure$.

An \emph{invariant} $I$ for  $D=\langle S,\ourmapsto\rangle$ is a subset $I\subseteq S$
such that
$$s_0\in I\land s_0\ourmapsto s_1\quad\Rightarrow\quad s_1\in I.$$
We write a fact $s\in I$ as $I(s)$ and refer to $s$ as an \emph{$I$ state}.
The \emph{restriction of $D$ to $I$} is the transition system $\langle I, \stackrel I\ourmapsto\rangle$
where $\stackrel I\ourmapsto$ is the restriction of $\ourmapsto$ to 
$I$.
A set of \emph{allowed initial states} $S'\subseteq S$ defines an invariant of those states \emph{reachable}
from some $s\in S'$, i.e., $\text{reachable}(S')=\{s'\mid s\in S' \land s\mapstostar s'\}$.
A \emph{(state) equivalence} is an equivalence  relation over $S$,
typically denoted
$\approx$. In the context of an invariant $I$, the relations $\approx$ and $\ourmapsto$ are understood to be restricted to $I$.

The
following 
$\alpha$ and $\beta$ corners\footnote{In recent literature within term rewriting, the terms peaks and cliffs have been
used for $\alpha$ and $\beta$ corners, respectively.} were
introduced in~\cite{DBLP:conf/lopstr/ChristiansenK14,DBLP:journals/fac/ChristiansenK17},
being implicit in~\cite{DBLP:journals/jacm/Huet80}.
An \emph{$\alpha$ corner}
is a structure $s_1\mapsfrom s_0\ourmapsto s_2$, where $s_0, s_1, s_2\in S$
and the indicated relationships hold;
$s_0$ is called a \emph{common ancestor}
and $s_1, s_2$ \emph{wing} states.
A \emph{$\beta$ corner}
is a structure $s_1\approx s_0\ourmapsto s_2$, where $s_0, s_1, s_2\in S$
and the indicated relationships hold.
In the context of an invariant $I$, the different types of corners are  defined only for $I$ states. 

Two  states  $s_1,s_2$  are \emph{joinable (modulo $\approx$)} whenever there exist paths
$s_1 \mapstostar s_1'$ and $s_2 \mapstostar s_2'$ with
$s_1'= s_2'$ ($s_1'\approx s_2'$).
A corner $s_1\mathrel{\mathit{Rel}} s_0\ourmapsto s_2$ is \emph{joinable (modulo $\approx$)} when 
$s_1,s_2$ are  joinable (modulo $\approx$); $\mathit{Rel}\in\{{\mapsfrom},{\approx}\}$. 

A transition system $D=\langle S,\ourmapsto\rangle$ is
\emph{confluent (modulo $\approx$)} whenever
$$
s_1\mapsfromstar s_0\mapstostar s_2 \quad\Rightarrow\quad \mbox{$s_1$ and $s_2$ are joinable (modulo $\approx$).}
$$
%
%
It is \emph{locally confluent} (modulo equivalence $\approx$) whenever
all its $\alpha$ ($\alpha$ and $\beta$) corners are joinable.
The following properties are fundamental.

\begin{lemma}[Newman \cite{Newman42}]\label{lm:newman}
A terminating transition system (under invariant $I$) is confluent if and only if
it is locally confluent.
\end{lemma}

\begin{lemma}[Huet \cite{DBLP:journals/jacm/Huet80}]\label{lm:huet-mod-eq}
A terminating transition system (under invariant $I$) is confluent modulo $\approx$ if and only if
it is locally confluent modulo $\approx$.
\end{lemma}
These properties reduce proofs of confluence (mod.\ equiv.) for terminating systems to proofs of the simpler property
of local confluence (mod.\ equiv.), but still, this may leave an infinite number of corners to be examined.
%
%
%

\section{Abstract Simulation}\label{sec:simulation}
Consider two transition systems,
$D_O=\langle S_O, \rightarrowsub{O}\rangle$ and $D^M=\langle S^M, \rightarrowsup{M}\rangle$, 
referred to as \emph{object} and \emph{meta level} systems.
A \emph{replacement} is a (perhaps partial) function $\rho\colon S^M \rightarrow S_O$; the application of $\rho$ to
some $s\in S^M$ is written $s\rho$.
For any structure $f(s_1,\ldots s_n)$  with states $s_1,\ldots s_n$ of $D^M$ (a transition, a tuple, etc.),
replacements apply in a compositional way,
$f(s_1,\ldots s_n)\rho = f(s_1\rho,\ldots s_n\rho).$
For a family of replacements $P=\{\rho_i\}_{i\in\mathit{Inx}}$, the \emph{covering} (or \emph{concretization})
of a structure $f(s_1,\ldots s_n)$ is defined as
\begin{eqnarray*}
\denotes {f(s_1,\ldots s_n)}_O^M & =& \{f(s_1,\ldots s_n)\rho\mid \rho\in P\}.
\end{eqnarray*}
Notice that $P$ is left implicit in this notation, as in the context of given object and meta level systems, there will be one and only one 
replacement family.

\begin{definition}\label{def:simulation}
An \emph{abstract simulation of $D_O$ by $D^M$} with possible invariants $I_O$, resp., $I^M$,
and equivalences $\approxsub O$, resp., $\approxsup M$, is defined by a family of replacements
$P=\{\rho_i\}_{i\in\mathit{Inx}}$ which satisfies the following conditions. 
\begin{eqnarray*}
s_0\rightarrowsup{M} s_1
   &  \quad\Rightarrow\quad  &
   \forall\rho\in P\colon\; s_0\rho\rightarrowsub{O} s_1\rho \;\lor\; s_0\rho=s_1\rho
\\
I^M(s)
   &  \quad\Rightarrow\quad  &
   \forall\rho\in P\colon\; I_O(s\rho)
\\
s_0\approxsup{M} s_1
   &  \quad\Rightarrow\quad  &
   \forall\rho\in P\colon\; s_0\rho\approxsub{O} s_1\rho
\end{eqnarray*}
\end{definition}
Notice that an abstract simulation does not necessarily cover all object level states, transitions, etc.
\begin{example}\label{ex:simulation}
Let $A=\{a_1, a_2, \ldots\}$, $B=\{b_1, b_2, \ldots\}$ and $C=\{c_1, c_2, \ldots\}$ be sets of states,
and $O$ and $M$ the following transition systems.
\begin{eqnarray*}
O & \quad=\quad & 
    \langle A\cup B\cup C, \{a_i\rightarrowsub{O} b_i \mid i=1,2,\ldots\}\cup\{a_i\rightarrowsub{O} c_i \mid i=1,2,\ldots\}\rangle \\
M & \quad=\quad & \langle \{a,b,c\},
   \{a\rightarrowsub{M} b,a\rightarrowsub{M} c\}\rangle 
\end{eqnarray*}
Assume equivalences $b\approxsup M\!\! c$ and $b_i\approxsub O\! c_i$, for all $i$.
Then the family of replacements $P=\{\rho_i\}_{i=1,2,\ldots}$, where $a\rho_i= a_i$, $b\rho_i =b_i$ and $c\rho_i= c_i$,
defines a simulation of $O$ by $M$.
It appears that $O$ and $M$ are not confluent, 
cf.\ the non-joinable corners $b_1\mapsfrom_{O} a_1\rightarrowsub{O} c_1$ and  $b\mapsfrom_{M} a\rightarrowsub{M} c$,
but both are confluent
modulo $\approxsub O$ ($\approxsup M$).
\end{example}
A meta level structure $m$ \emph{covers} an object structure $k$ whenever $k\in\denotes{m}_O^M$.
When $\denotes{m}_O^M=\emptyset$, $m$ is \emph{inconsistent.}
When $\denotes{m'}_O^M\subseteq \denotes{m}_O^M$, $m'$ is a \emph{substate}/\emph{subcorner}, etc.\ of $m$,
depending on the inherent type of $m$.
%
When $D_O$ and $D^M$ both include $\failure$,
it is required that
$\denotes{\failure}_O^M = \{\failure\}$. 
A given meta level state $S$ is \emph{mixed} whenever $\denotes{S}_O^M$ includes
both $\failure$ and non-$\failure$ states.
Transitions are only allowed from consistent and neither failed nor mixed states.

The following  is a consequence of the definitions. 
\begin{lemma}\label{lm:simulation-and-concluence}
An object level corner, which is covered by a joinable (mod.\ equiv.) meta level corner, is joinable (mod.\ equiv.).
\end{lemma}
When doing confluence proofs, we may search for a small set of
\emph{critical} meta level corners,\footnote{In the literature, the term \emph{critical pair} is used for the
pair of wing states of our critical corners. }
whose joinability guarantees joinability of any object level corner, i.e.,
any other object level corner not covered by one of these is seen to be joinable in other ways.
For term rewriting systems, e.g.,~\cite{BaderNipkow1999}, and previous work on CHR, such critical sets have been defined by explicit constructions.

We introduce a mechanism for splitting a meta level  corner  $\Lambda$ into a set
of  corners, which together covers the same set of object corners as $\Lambda$.
This is useful when  $\Lambda$ in itself is not joinable, but each of the new corners are.
In some cases, splitting is  necessary for proving confluence under an invariant as shown in Section~\ref{sec:conflLogicBased}  and exemplified in Examples~\ref{ex:zigzagPlain} and~\ref{ex:zigzagPlainCont}.

\begin{definition}
Let $s$ be a  meta level state (or corner).
A set of states (or corners)
$\{ s_i\}_{i\in\mathit{Inx}}$ is a \emph{splitting} of $s$ whenever $\bigcup_{i\in\mathit{Inx}}\denotes{ s_i}_O^M = \denotes{ s\,}_O^M$.
A corner (set of corners) is \emph{split joinable} (mod.\ equiv.) if it (each of its corners) is joinable (mod.\ equiv.), inconsistent, or has a splitting into a set of split joinable (mod.\ equiv.) corners.
\end{definition}
\begin{corollary}\label{cl:simulation-and-concluence}
An object level corner, which is covered by a split joinable (mod.\ equiv.) meta level corner, is joinable (mod.\ equiv.).
\end{corollary}
\section{Constraint Handling Rules}\label{sec:chr}
%
Most actual implementations of CHR are fully deterministic, i.e., for a given query, there is at most one answer
state (alternatively, the program is non-terminating). In this light, it may be discussed whether confluence is an interesting property,
and if so, to what extent the applied semantics
should be nondeterministic.
Our thesis is the following: choice of next constraints to be tried and which rule to be used should
be nondeterministic.
Thus a confluent program can be understood by the programmer without considering the
detailed control mechanisms in the used implementation; this also anticipates parallel implementations.
We see only little interest in considering confluence for the so-called refined CHR semantics~\cite{DuckSBH04}
in which only very little nondeterminism is retained.
%
%

Similarly to~\cite{DBLP:conf/lopstr/ChristiansenK14,DBLP:journals/fac/ChristiansenK17}, we remove w.l.o.g.\ two redundancies from the logic-based semantics~\cite{DBLP:conf/cp/Abdennadher97,fru-chr-book-2009}: global variables and the two-component constraint store.
\begin{itemize}
  \item Global variables are those in the original query.
  Traditionally they are kept as a separate state-component, such that values bound to them
  can be reported to the user at the end. 
  The same effect can be obtained by a constraint
  \texttt{global}/2 that does not appear in any rule, but may be used in the original query:
  writing \texttt{?-} \texttt{p(X)} as \texttt{?-} \texttt{p(X),} \texttt{global('X',X)}, means that the value of the variable named \texttt{'X'} can be read out as
  the second argument of this constraint in a final state.
  \item We avoid separating the constraint store into query and active parts, 
  as the transition sequences with or without this
  separation are essentially the same.
\end{itemize}

\subsection{Syntax}\label{sec:syntax}
Standard first-order notions of variables, terms, predicates atoms, etc.\ are assumed.
Two disjoint sets of \emph{constraint predicates}
are assumed, \emph{user constraints} and \emph{built-in constraints};
the actual set of built-ins may vary depending on the application.
We use the generalized simpagation form~\cite{fru-chr-book-2009} to capture all rules of CHR.
A \emph{rule} is a structure of the form
$$H_1 \texttt{\char92}H_2\;\; \texttt{<=>} \;\;G \texttt{|} C$$
where $H_1 \texttt{\char92}H_2$ is the \emph{head} of the rule,
 $H_1$ and $H_2$ being sequences, not both empty, of user constraints;
$G$ is the \emph{guard} which is a conjunction of built-in constraints; 
and $C$ is the \emph{body} which is a sequence of constraints of either sort.
When $H_2$ is empty, the rule is a \emph{simplification}, which may be written
$H_1$ \texttt{<=>} $G \texttt{|} C$;
when $H_2$ is empty, it is a \emph{propagation}, which may be written $H_2$ \texttt{==>} $G \texttt{|} C$;
any other rule is a \emph{simpagation}; when $G=\true$,  $(G \texttt{|})$ may be left out. 
The \emph{head variables} of a rule are those appearing in the head, any other variable is \emph{local}.
%
%
The following notion is convenient when defining the CHR semantics and its meta level simulation.

\begin{definition}
A \emph{pre-application} of a rule $r = (\nonamerule{H_1}{H_2}GC)$ is of the form\break
$(\nonamerule{H_1'}{H_2'}{G'}{C'})\sigma$
where 
$r' = (\nonamerule{H_1'}{H_2'}{G'}{C'})$ is a variant of $r$ with fresh variables 
and $\sigma$ is a substitution to the head variables of $r'$, where, for no variable $x$, $x\sigma$ contains a local variable of $r'$.
\end{definition}
%
%
The operator $\uplus$ refers to union of multisets, so that, e.g.,
$\{a,a\}\uplus\{a\}=\{a,a,a\}$; 
for difference of multisets, we use standard notation for set difference, assuming it
takes into account the number of copies, e.g., $\{a,a\}\setminus\{a\}=\{a\}$.


\subsection{The logic-based operational semantics for CHR}\label{sec:logic-semantics}
The semantics presented here is essentially identical to the one used by~\cite{DBLP:conf/cp/Abdennadher97}
and the so-called abstract operational semantics $\omega_t$ of~\cite{fru-chr-book-2009},
taking into account the simplifications explained above.
Following~\cite{RaiserEtAl2009}, we define a state as  an equivalence class, abstracting away the specific variables used
and the different ways  the same logical meaning can be expressed by different 
conjunctions of built-ins.\footnote{Raiser et al~\cite{RaiserEtAl2009} defined ``state'' similarly to what we call state representation, and
they defined an operational semantics over equivalence classes of such states.
We have taken the natural step of promoting such equivalence classes to be our states.}
A logical theory $\mathcal{B}$ is assumed for the built-in predicates.

A \emph{state representation (s.repr.)} is a pair $\langle S, B\rangle$,
where  the \emph{constraint store} $S$  is a multiset of constraint atoms 
and the \emph{built-in store} $B$ is a conjunction of built-ins;
any s.repr.\ with an unsatisfiable built-in store is considered identical to $\failure$.
Two s.repr.s $\langle S, B\rangle$ and $\langle S', B'\rangle$ are \emph{variants}  whenever, either\footnote{An equation
between multisets should be understood as an equation between suitable permutations of their elements.}
\begin{itemize}
  \item they are both $\failure$, or
  \item there is a renaming substitution $\rho$ such that\\
  $\mathcal{B}\models\, \forall(B\rho\rightarrow\exists(S\rho=S'\land B'))
  \, \land\,  \mathcal{B}\models \forall(B'\rightarrow\exists(S\rho=S'\land B\rho))$
\end{itemize}
A \emph{state} is an equivalence class of s.repr.s under the variant relationship.
For simplicity of notation, we typically indicate a state by one of its s.repr.s.

%
%
A \emph{rule application} w.r.t.\ to a non-failure state $\langle S, B\rangle$
is a pre-application $H_1 \texttt{\char92}H_2\;\texttt{<=>} \;G \texttt{|} C$
for which $\mathcal{B}\models B\rightarrow\exists_L G$, where $L$ is the list of its local variables.
There are two sorts of transitions, \emph{by rule application} and \emph{by built-in}.
%
%
\begin{eqnarray*}
\langle H_1\uplus H_2\uplus S, B\rangle & \;\rightarrowsublogic\; & \langle H_1\uplus C\uplus S, G\land B\rangle\\
&&  \text{when there exists a rule application $H_1 \texttt{\char92}H_2 \texttt{<=>} G \texttt{|} C$}, \\
\langle \{b\}\uplus S, B\rangle & \;\rightarrowsublogic\; & \langle  S, b\land B\rangle\qquad \text{for a built-in $b$}.
\end{eqnarray*}

\section{Confluence under the logic-based semantics re-explained, and why invariants are difficult}\label{sec:conflLogicBased}
Here we explain the results of~\cite{DBLP:conf/cp/Abdennadher97,DBLP:conf/cp/AbdennadherFM96},
also summarized in~\cite{fru-chr-book-2009}, using abstract simulation.
Object and meta level systems coincide and are given by a CHR program under the logic-based semantics.
%
Two rules give rise to a critical corner if a state can be constructed in which one rule consumes
constraints that the other one needs
to be applied; in that case, rule applications do not commute
and a specific proof of joinability must be considered.
We anticipate the re-use of the construction,
when invariants are introduced: in a \emph{pre}-corner, the guards are not necessarily satisfied
(but may be so in the context of an invariant).

\begin{definition}\label{def:criticalClassic}
Consider two rules
$\namedrule r{H_1}{H_2}GC$ and
$\namedrule {r'}{H'_1}{H'_2}{G'}{C'}$ 
renamed apart,
and let $A$ and $A'$ be non-empty sets of constraints such that
  $A\subseteq H_2$, $A'\subseteq H'_1\uplus H'_2$
and $\mathcal{B}\models \exists(A{=}A')$.
%
In that case, let 
\begin{eqnarray*}
\bar H &=& (H_1{\uplus} H_2 {\uplus} H'_1{\uplus} H'_2)\setminus A \\
s_0&=& \langle \bar H, ( G{\land} G'{\land} A{=}A')\rangle\\
s&=&\langle \bar H{\setminus}H_2{\uplus} C, (G{\land} G'{\land}A{=}A')\rangle\\
s'&=&\langle \bar H{\setminus}H'_2{\uplus} C', (G{\land}G{'\land}A{=}A')\rangle\\
\end{eqnarray*}
When $s\neq s'$, $s_0$ is a \emph{critical, common ancestor state}, and 
$s\leftarrowsublogic s_0\rightarrowsublogic s'$ is a \emph{critical $\alpha$ pre-corner};
the constraints $A$ (or $A'$) is called the \emph{overlap} of $r$ and $r'$.
When, furthermore, $\mathcal{B}\models \exists(G\land G'\land A{=}A')$, it is  a
\emph{critical} $\alpha$ corner.
\end{definition}
The simulation is given by the following cover function.
\begin{eqnarray*}
\denoteslogiclogic{\langle S,B\rangle} & = & \{ \langle S\uplus S^+,B\land B^+\rangle  \mid\\
&& \hbox to 0.4em{}\text{$S^+$ is a multiset of user and built-in constraints,}\\
&& \hbox to 0.4em{}\text{$B^+$ is a conjunction of built-ins} \, \}
\\
\denoteslogiclogic{\langle S,B\rangle\!\rightarrowsublogic\!\langle S',B'\rangle} & = & 
   \{ (\langle S\uplus S^+,B\land B^+\rangle\rightarrowsublogic\langle S'\uplus S^+,B'\land B^+\rangle)  \mid
   \\&&\hbox to 0.4em{} \text{$S^+$ is a multiset of user and built-in constraints,}
  \\&&\hbox to 0.4em{} \text{$B^+\!$ is a conjunction of built-ins, $\exists( B{\land} B^+)$ holds}  \}
\end{eqnarray*}
%
It is easy to check that this definition satisfies the conditions for being an abstract simulation given in Section~\ref{sec:simulation},
relying on \emph{monotonicity}: $\mathcal{B}\models B\land B^+\rightarrow\exists_L G$\label{intext:monotonicityproperty}.

It can be shown that any corner not covered by a critical corner (Definition~\ref{def:criticalClassic})
is trivially joinable, see the extended report~\cite{ConfModEqforCHRRevisited}.
%
Thus, according to Lemmas~\ref{lm:newman} and~\ref{lm:simulation-and-concluence},
the program under investigation is confluent whenever it is terminating
and this set of
critical corners is joinable.
The set of critical corners is finite and that 
 allows for automatic confluence proofs by checking the critical corners, one by one, e.g.,~\cite{Raiser-Langbein2010}.

\begin{example}\label{ex:set-critical-corners-logic}
Consider the one-rule \texttt{set}-program of Example~\ref{ex:set-informal}, ignoring invariant
and state equivalence.
There are two critical corners, given by the two ways, the rule can overlap with itself:

\tikzset{|/.tip={Bar[width=.8ex,round]}}
\noindent\begin{tikzpicture}[|->, auto,node distance=2cm and 1.0cm, 
               main node/.style={font=\normalsize 
              }]
  \node[main node] (1) {$\langle\{\texttt{item(X1)},\texttt{set(L)},\texttt{item(X2)}\}, \true\rangle$};
  \node[main node] (2) [above =0.3cm of 1] {$\langle \{\texttt{set([X1|L])},\texttt{item(X2)}\}, \true\rangle $};
  \node[main node] (3) [below =0.3cm of 1] {$\langle \{\texttt{item(X1)},\texttt{set([X2|L])}\}, \true\rangle$};
  \path
    (1) edge [shorten >=-1pt, shorten <=0pt] node [near start,right] {\scriptsize $logic$} (2)
    (1) edge [shorten >=-1pt, shorten <=0pt] node [near end,right] {\scriptsize $logic$} (3);

\begin{scope}[xshift = 6.3 cm]
  \node[main node] (1) {$\langle\{\texttt{set(L1)},\texttt{item(X)},\texttt{set(L2)}\}, \true\rangle$};
  \node[main node] (2) [above =0.3cm of 1] {$\langle\{\texttt{set([X|L1])},\texttt{set(L2)}\}, \true\rangle$};
  \node[main node] (3) [below =0.3cm of 1] {$\langle\{\texttt{set(L1)},\texttt{set([X|L2])}\}, \true\rangle$};
  \path
    (1) edge [shorten >=-1pt, shorten <=0pt] node [near start,right] {\scriptsize $logic$} (2)
    (1) edge [shorten >=-1pt, shorten <=0pt] node [near end,right] {\scriptsize $logic$} (3);
\end{scope}
\end{tikzpicture}
None of these corners are  joinable, so the program is not confluent.
\end{example}
The simulation defined above, 
relying on monotonicity, 
do not generalize well for confluence under invariant, referred to as ``observable confluence'' in~\cite{DBLP:conf/iclp/DuckSS07}.
\begin{example}\label{ex:zigzagPlain}
Consider the CHR program consisting of the following four rules.
\begin{center}
\begin{tabular}{ccl}
$r_1$:\quad\texttt{p(X)} \texttt{<=>} \texttt{q(X)} &\qquad\qquad &
   $r_3$:\quad\texttt{q(X)} \texttt{<=>} \texttt{X>0 |} \texttt{r(X)} \\
$r_2$:\quad\texttt{p(X)} \texttt{<=>} \texttt{r(X)} &\qquad\qquad &
   $r_4$:\quad\texttt{r(X)} \texttt{<=>} \texttt{X$\ttleq$0 |} \texttt{q(X)}
\end{tabular}
\end{center}
It is not confluent as its single critical corner
$\texttt{q(X)}\,{\mapsfrom}\,\texttt{p(X)}\,{\ourmapsto}\,  \texttt{r(X)}$
is not joinable (the built-in stores are $\true$ and thus omitted).
However, adding the invariant ``reachable from an initial state \texttt{p($n$)} where $n$ is an integer''
makes it confluent. We indicate the set of all non-trivial object level corners as follows, with the dashed transitions
proving each of them joinable. 
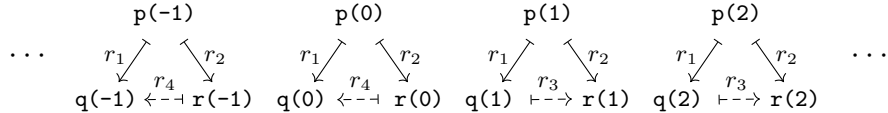
\begin{center}
\tikzset{|/.tip={Bar[width=.8ex,round]}}
\begin{tikzpicture}[|->, auto,node distance=2cm and 1.0cm,
               main node/.style={font=\small,text width={width("p(-1)")+2pt}, text centered}]

  \node[main node] (1) {\texttt{p(0)}};
  \node[main node] (2) [below left=0.6cm and -0.3cm of 1] {\texttt{q(0)}};
  \node[main node] (3) [below right=0.6cm and -0.3cm of 1] {\texttt{r(0)}};

  \path
    (1) edge [shorten >=-1pt, shorten <=3pt] node [midway,left] {\small$r_1$} (2)
    (1) edge [shorten >=-1pt, shorten <=3pt] node [midway,right] {\small$r_2$} (3)
    (3) edge [densely dashed, shorten >=-2pt, shorten <=0pt] node [above] {\small$r_4$} (2);  
    
  \begin{scope}[xshift = 2.5 cm]
  \node[main node] (1) {\texttt{p(1)}};
  \node[main node] (2) [
  below left=0.6cm and -0.3cm of 1] {\texttt{q(1)}};
  \node[main node] (3) [below right=0.6cm and -0.3cm of 1] {\texttt{r(1)}};

  \path
    (1) edge [shorten >=-1pt, shorten <=3pt] node [midway,left] {\small$r_1$} (2)
    (1) edge [shorten >=-1pt, shorten <=3pt] node [midway,right] {\small$r_2$} (3)
    (2) edge [densely dashed, shorten >=-2pt, shorten <=0pt] node [above] {\small$r_3$} (3);  

  \end{scope}

  \begin{scope}[xshift = 5 cm]
  \node[main node] (1) {\texttt{p(2)}};
  \node[main node] (2) [
  below left=0.6cm and -0.3cm of 1] {\texttt{q(2)}};
  \node[main node] (3) [below right=0.6cm and -0.3cm of 1] {\texttt{r(2)}};

  \path
    (1) edge [shorten >=-1pt, shorten <=3pt] node [midway,left] {\small$r_1$} (2)
    (1) edge [shorten >=-1pt, shorten <=3pt] node [midway,right] {\small$r_2$} (3)
    (2) edge [densely dashed, shorten >=-2pt, shorten <=0pt] node [above] {\small$r_3$} (3);  

  \end{scope}
  
  \begin{scope}[xshift = -4.4 cm, yshift= -0.55cm]
  \node[main node] (1) {\large$\cdots$};
    \end{scope}
    
    \begin{scope}[xshift = 6.8 cm, yshift= -0.55cm]
  \node[main node] (1) {\large$\cdots$};
    \end{scope}  
      
\begin{scope}[xshift = -2.6 cm]
  \node[main node] (1) {\texttt{p(-1)}};
  \node[main node] (2) [
  below left=0.6cm and -0.3cm of 1] {\texttt{q(-1)}};
  \node[main node] (3) [below right=0.6cm and -0.3cm of 1] {\texttt{r(-1)}};

  \path
    (1) edge [shorten >=-1pt, shorten <=3pt] node [midway,left] {\small$r_1$} (2)
    (1) edge [shorten >=-1pt, shorten <=3pt] node [midway,right] {\small$r_2$} (3)
    (3) edge [densely dashed, shorten >=-1pt, shorten <=0pt] node [above] {\small$r_4$} (2);  

  \end{scope}      
\end{tikzpicture}
\end{center}
%
These object corners and their proofs of joinability obviously fall in two groups of similar shapes, but there is no
way to construct a finite set (of, say, one or two elements) that covers all object corners.
In other words, the smallest set of meta level corners that covers this set is the set itself.
This was also noticed in~\cite{DBLP:conf/iclp/DuckSS07} that used a construction that essentially reduces to the abstract simulation shown above.
\end{example}
The abstract simulation given by $\denoteslogiclogic-$ of Definition~\ref{def:criticalClassic} above
defines an abstract interpretation, whose abstract domain is the complete lattice of CHR states ordered by the substate relationship (Section~\ref{sec:simulation}).
Referring to Example~\ref{ex:zigzagPlain}, for instance the join of the infinite set of states $\{\langle\texttt{p($t$)}, b\rangle\mid t$ is a term, $b$ is a  conjunction of built-ins$\}$
is $\langle\texttt{p(X)},\true\rangle$.
When the grounding invariant is introduced, the join operator is not complete; an attempt to join, say, $\langle\texttt{p(0)},\true\rangle$ and
$\langle\texttt{p(1)},\true\rangle$ would not satisfy the invariant.\footnote{Such an attempt might be $\langle\texttt{p(X)},(\texttt{X=0}\lor\texttt{X=1})\rangle$;
notice that \texttt{X} is a variable, thus breaking the invariant.}

\section{Invariants and modulo equivalence}\label{sec:inf-mod-eq}
A program is typically developed with an intended set of queries in mind, giving rise to
a state invariant, which may make an otherwise non-confluent program observably confluent  (mod.\ equiv.).
We can indicate a few general patterns of invariants and their possible effect on  confluence.
\begin{itemize}
  \item Elimination of non-joinable critical corners that do not cover any object corner satisfying the invariant.
  This was shown in Example~\ref{ex:zigzagPlain} above,
  and is also demonstrated in the continuation of Example~\ref{ex:set-critical-corners-logic} (Ex.~\ref{ex:set-critical-corners-logicCont}, below): ``only one \texttt{set} constraint allowed".
  \item Making it possible to apply a given rule, which otherwise could not apply, e.g.,
  providing a ``missing''  head constraint or enforcing guard satisfaction:
  \begin{enumerate}
\item $\!$``if a state contains  \texttt{p(}\scalebox{.9}[1.0]{\textit{something}}\texttt{)},
               it also contains  \texttt{q(}\scalebox{.9}[1.0]{\textit{the-same-something}}\texttt{)}'',
     \item $\!$``if a state contains  \texttt{p(}\scalebox{.9}[1.0]{\textit{something}}\texttt{)}, this \scalebox{.9}[1.0]{\textit{something}} is a constant $>1$''.
   \end{enumerate}
\end{itemize}
An invariant of type 1 ensures confluence mod.\ equiv. of a version of the Viterbi algorithm \cite{DBLP:journals/fac/ChristiansenK17}; an invariant of type 2 is indicated in Example~\ref{ex:zigzagPlain} 
  and formalized in Example~\ref{ex:zigzagPlainCont}, below.

As shown in Example~\ref{ex:zigzagPlain} above, invariants block for a direct re-use CHR's logical semantics as
its own meta-level and, accordingly, existing methods and confluence checkers.
In some cases, it is possible to eliminate invariants by program transformations, so that rules apply exactly
when the invariant and the original rule guards are satisfied; this means that the transformed program is confluent if and only if the original one is
confluent under the invariant.
\begin{example}\label{ex:transformationalapproach}
 Reconsidering the program of Example~\ref{ex:zigzagPlain}, the following is an example of such a transformed program; the constants \texttt{a} and \texttt{b} are introduced as representations of positive, resp., non-positive integers.
\begin{verbatim}
   p(a) <=> q(a).    p(a) <=> r(a).    p(a) <=> r(a).
   p(b) <=> q(b).    p(b) <=> r(b).    r(b) <=> q(b).
\end{verbatim}
\end{example}
Such program transformations become more complex when the guards describe more involved dependencies between the head variables.
More importantly, invariants that exclude certain constraints in a state cannot be expressed in this way,
for example ``only one \texttt{set} constraint allowed" (Examples~\ref{ex:set-critical-corners-logic} and~\ref{ex:set-critical-corners-logicCont}). Thus we refrain from pursuing a transformational approach.
%
%
To obtain a maximum degree of generality,
we introduce a meta level formalization of CHR's operational semantics that include representations
as explicit data objects of states and their components, possibly parameterized by constrained meta variables.

\subsection{The choice of a ground representation}
Invariants and state equivalences are inherently meta level statements, as they are \emph{about} states,
and may refer to notions inexpressible at the object level, 
e.g., that some part being ground or a variable.
Earlier work on meta-interpreters for  logic programs, e.g.,~\cite{DBLP:journals/jlp/Christiansen98,Hill94meta-programming-in-logic,DBLP:conf/meta/HillL88},  offers the desired expressibility in terms of a 
\emph{ground representation}.
Any object term, formula, etc.~is named by a ground meta level term.
Variables are named by special constants, say \texttt{X} by \texttt{'X'},
and any other symbol by a function symbol written the same way;
e.g., the non-ground object level atom \texttt{p(A)} is named by the ground meta level term \texttt{p('A')}.
For any such ground meta level term $mt$, we indicate the object it names as $\denotesgr{mt}$. For example,
  $\denotesgr{\texttt{p('A')}}= \texttt{p(A)}$ and
  $\denotesgr{\texttt{p('A')}\land\texttt{'A'>2}} = (\texttt{p(A)}\land\texttt{A>2})$.

For a given object entity $e$, we define its \emph{lifting}\label{inline:lifting}
 to the meta level by
1) selecting a meta level term that names $e$, and 2) replacing variable names
in it consistently by fresh meta level variables.
For example, $\texttt{p(X)}\land\texttt{X>2}$ is lifted to $\texttt{p($x$)}\land\texttt{$x$>2}$, where \texttt{X} and $x$
are object, resp., meta variables.
By virtue of this overloaded syntax, we may read  such an entity $e$ (implicitly) as its lifting.


A collection of \emph{meta level constraints} is assumed whose meanings are given by a theory
$\Mmetalogic$.
We start defining meta level states without detailed assumptions about $\Mmetalogic$, that are postponed
to Definition~\ref{def:meta-logic-theory} below.
We assume object level built-in theory $\mathcal{B}$, invariant $I\sublogic$ and state equivalence $\approxlogic$.

\begin{definition}\label{def:meta-logic-state}
A \emph{constrained meta level term} is a structure of the form\break$(mt \where M)$, where $mt$ is a meta level term and $M$
a conjunction of  $\Mmetalogic$ constraints.
We define
\begin{eqnarray*}
[M]&\quad=\quad& \{ \sigma \mid \Mmetalogic\models M\sigma\},\\
\denotesmetalogic{mt\where M}&\quad=\quad& \{ \denotesgr{mt\,\sigma} \mid \sigma \in [M]\}.
\end{eqnarray*}
A \emph{meta level state representation (s.repr.)} is a constrained meta level term $st \where M$ for which
$\denotesmetalogic{st \where M}$ is a set of object level states.
Two meta level s.repr.s $SR_1$, $SR_2$ are \emph{variants} whenever each
object level s.repr.\ in $\denotesmetalogic{SR_1}$ is a variant of some object level s.repr.\ in $\denotesmetalogic{SR_2}$
and vice versa. A \emph{meta level state}
is an equivalence class of meta level s.repr.s under the variant relationship.
%
%
For structures of meta level states (transitions, corners, etc.),
we apply the following convention, where $f$ may represent any such structure.
\begin{eqnarray*}
  & & \denotesmetalogic{f(mt_1\where M_1, \ldots, mt_n\where M_n)} \\
 & =\quad & \denotesmetalogic{f(mt_1, \ldots, mt_n) \where M_1 \land \ldots \land M_n} 
\end{eqnarray*}
Meta level invariant $\Imetalogic$ and equivalence $\approxmetalogic$ are defined as follows.
\begin{itemize}
  \item $\Imetalogic(S)$ whenever $I\sublogic(s)$  for all $s\in\denotesmetalogic S$.
  \item $S_1 \approxmetalogic S_2$ whenever $s_1 \approxlogic s_2$  for all $(s_1,s_2)\in \denotesmetalogic{(S_1,S_2)}$. 
\end{itemize}
\end{definition}
As before, we may indicate a meta level state by a representation of it.
%

\begin{definition}\label{def:meta-logic-theory}
The theory $\Mmetalogic$ includes at least the following constraints.
\begin{itemize}
  \item {\rm\texttt=}\emph{/2} with its usual meaning of syntactic identity,
  \item \emph{Type constraints} {\rm\texttt{type}}\emph{/2}. For example {\rm\texttt{type(var,$x$)}} is  true in $\Mmetalogic$ whenever $x$ is the name of an object level variable; {\rm\texttt{var}} is an example of a \emph{type}, and we introduce more types below when we need them.
  \item \emph{Modal constraints} $\willsucced F$ and $\willfail F$
     defined to be true in $\Mmetalogic$ whenever  $\mathcal{B}\models\denotesgr{F}$, resp.,
     $\mathcal{B}\models\denotesgr{\lnot F}$.
  \item We define two constraints $\cinv$ and $\cequiv$
  such that $\cinv(\Sigma)$ is true in $\Mmetalogic$ whenever $\denotesgr{\Sigma}$ is an $\Ilogic$ state (representation) of the logical semantics,
  and $\cequiv(\Sigma_1,\Sigma_2)$ whenever  $\denotesgr{(\Sigma_1,\Sigma_2)}$ is a pair of
  states  (representations) $(s_1,s_2)$ of the logical semantics such that  $s_1\approxlogic s_2$.
  \item {\rm\texttt{freshVars($L$,$T$)}}  is true in $\Mmetalogic$ whenever $L$ is a list of all different variables names, none of which occur in the term $T$; {\rm\texttt{freshVars($L_1$,$L_2$,$T$)}} ab\-bre\-vi\-a\-tes
  {\rm\texttt{freshVars($L_{12}$,$T$))}} where $L_{12}$ is the concat.\ of $L_1$ and $L_2$.
\end{itemize}
\end{definition}
%
Definitions~\ref{def:meta-logic-state} and~\ref{def:meta-logic-theory} comprise the first steps
towards a simulation of the logic-based semantics, and  we continue with the last part,
the transition relation.

\begin{definition}\label{def:meta-logic-trans}
Consider a (lifted version of a) pre-application $H_1 \text{\rm\texttt{\char92}}H_2\;\text{\rm\texttt{<=>}} \;G \text{\rm\texttt{|}} C$
with local variables $L$
and a consistent meta level state $(S \where M)$ with $S=$\break $\langle H_1{\uplus} H_2{\uplus} S^+,B^+\rangle$
and 
$$\Mmetalogic\quad\models\quad M\;\rightarrow\; \bigl( \cinv(S)\land\willsucced B^+\land\willsucced(B^+{\rightarrow}\exists_LG)
\land  \text{\rm\texttt{freshVars($L$,$S$)}}\bigr).
$$
Then the following is a \emph{meta level transition by rule application}.
$$S \where M\
\quad\ourlongmapstometalogic\quad
\langle H_1{\uplus} C{\uplus}S^+,G{\land} B^+\rangle \where  M$$
Consider a (lifted version of a) built-in $b$ of $\mathcal{B}$ and a consistent meta level state 
$(S \where M)$ with $S=\langle\{b\}{\uplus}S^+,  B^+\rangle$ and 
$$\Mmetalogic\quad\models\quad M\;\;\rightarrow\;\; \bigl( \cinv(S)\land\willsucced B^+\bigr).
$$
Then the following is a \emph{meta level transition by built-in}.
$$
\langle\{b\}{\uplus}S^+,B^+\rangle  \where M \quad\ourlongmapstometalogic\quad \langle S^+, b{\land}B^+\rangle  \where M
$$
\end{definition}
Notice that for both sorts of transitions, the implication of $\willsucced B^+$ excludes transitions from
failed and mixed states.
For built-in transitions, the resulting states may be non-failed, failed or mixed.

\begin{lemma}\label{lm:abs-sim-logic-based-sem}
For a given CHR program with $\Ilogic$ and $\approxlogic$, the definitions of
meta level states and transitions $\ourlongmapstometalogic$, $\Imetalogic$ and $\approxmetalogic$, together with $\denotesmetalogic-$
comprise an abstract simulation of the logic-based semantics.
\end{lemma}
%
Transitions are not possible from a mixed or failed meta level state,
but modal constraints are useful for restricting to the relevant substate, such that transitions are known to exists.
This is expressed by the following propositions that are immediate consequences of the definitions.
\begin{proposition}\label{prop:meta-rule-trans-substate}
Let  $r\colon H_1 \text{\rm\texttt{\char92}}H_2\; \text{\rm\texttt{<=>}} \;G  \text{\rm\texttt{|}} C$ be a
(lifted version of a) pre-application with local variables $L$ and  $\Sigma=(\langle S,B\rangle\where M)$
a meta level state with $H_1{\uplus}H_2\subseteq S$.
Whenever the meta level state $\Sigma^{\willsucced}=(\langle S,B\rangle\where M\land \widehat M)$
is consistent, with 
$\widehat M \;=\;  \cinv(\langle S,B\rangle)  \land \willsucced B \land \willsucced(B{\rightarrow}\exists_LG)
\land \text{\rm\texttt{freshVars($L$,$\Sigma$)}},$
there exists a meta level rule application by $r$,
$$\Sigma^{\willsucced}\quad\ourlongmapstometalogic\quad \langle S{\setminus} H_2{\uplus C},B{\land}G\rangle\where M\land \widehat M.$$
Furthermore, $\Sigma^{\willsucced}$ is the greatest substate of 
$\Sigma$
to which $r$ can apply.
\end{proposition}

\begin{proposition}\label{prop:meta-builtin-trans-substate}
Let $b$ be a
(lifted version of a) built-in and $\Sigma=(\langle S,B\rangle \where M)$
a meta level state with $b\in S$.
When $\Sigma^{\willsucced}=(\langle S,B\rangle\where M{\land} \widehat M)$
is cons- istent, with
 $\widehat{M^{\willsucced}} =  \cinv(\langle S,B\rangle)  {\land} \willsucced B {\land} \willsucced(B{\rightarrow}b),$
there is a meta level trans., 
$$\Sigma^{\willsucced}\quad\ourlongmapstometalogic\quad \langle S{\setminus}\{b\},B\land b\rangle  \where M \land \widehat{M^{\willsucced}}.
$$
Whenever $\Sigma^{\willfail}=(\langle S,B\rangle\where M\land \widehat{M^{\willfail}})$
is consistent, with
 $\widehat{M^{\willfail}}\;=\;  \cinv(\langle S,B\rangle)  \land \willsucced B \land \willfail(B{\rightarrow}b),$
there is a meta level transition by $b$,
$$\Sigma^{\willfail}\quad\ourlongmapstometalogic\quad \langle S{\setminus}\{b\},B\land b\rangle  \where M \land \widehat{M^{\willfail}}.
$$
The state $\Sigma^{\willsucced}$ (resp.\ $\Sigma^{\willfail}$)
is the greatest substate of $\Sigma$ for which the  meta level transition by $b$
leads to a non-failure and non-mixed (resp.\ failed) state.
\end{proposition}
With Propositions~\ref{prop:meta-rule-trans-substate}
and~\ref{prop:meta-builtin-trans-substate} in mind, we 
define meta level critical corners from the critical corners
of Definition~\ref{def:criticalClassic}.

\begin{definition}\label{def:critical-meta-level-corner:alpha}
Let $\langle S_1,B_1\rangle\leftarrowsublogic\langle S_0,B_0\rangle\rightarrowsublogic\langle S_2,B_2\rangle$
be a (lifted version of a) critical $\alpha$ pre-corner given by Def.~\ref{def:criticalClassic}, in which the leftmost (rightmost)
rule application has local variables $L_1$ ($L_2$) and guard $G_1$ ($G_2$).
Assume $S^+$ and $B^+$ are fresh meta level variables and
let, for $i=0,1,2$, 
\begin{eqnarray*}
\Sigma_i &\quad = \quad& \langle S_i{\uplus}S^+,B_i{\land}B^+\rangle \\
M &\quad =\quad & \cinv(\Sigma_0) \land \willsucced B_0 \land \willsucced(B_0{\land}B^+{\rightarrow}\exists_{L_1}G_1)\land\willsucced(B_0{\land}B^+{\rightarrow} \exists_{L_2}G_2)\land \\
&&  \text{\rm\texttt{freshVars($L_1$,$L_2$,$\Sigma$)}} 
\end{eqnarray*}
When $(\Sigma_0\where M)$ is consistent, the following is \emph{a critical meta level $\alpha$  corner}.
$$(\Sigma_1\where M)\;\;\ourlongmapsfrommetalogic\;\;(\Sigma_0\where M)\;\;\ourlongmapstometalogic\;\;(\Sigma_2\where M)$$
\end{definition}
\begin{example}(Continuing\ Ex.\ref{ex:zigzagPlain})
\label{ex:zigzagPlainCont}
The invariant is formalized at the meta level
as states of the form $\langle\{\mathit{pred}\texttt{($n$)}\},\true\rangle\where \texttt{type(int,$n$)}$ where $\mathit{pred}$ is one of
\texttt{p}, \texttt{q} and \texttt{r}.
Below is shown the non-joinable critical meta level corner generated by Def.~\ref{def:critical-meta-level-corner:alpha}. It is split-joinable as demonstrated by its splitting into two corners; each shown joinable
by the indicated dotted transition.
Let $M$ stand for the meta-level constraint $\texttt{type(int,$n$)}$, $M_1$ for  $M{\land}\willsucced n{\leq}0$ and $M_2$ for $M{\land} \willsucced n{>} 0$.
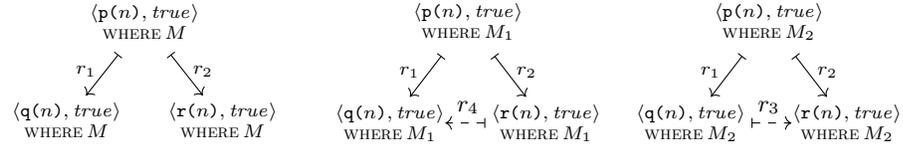
\begin{center}
\tikzset{|/.tip={Bar[width=.8ex,round]}}
\noindent\begin{tikzpicture}[|->, auto,node distance=2cm and 1.0cm,
               main node/.style={font=\scriptsize,
               text width={width("$<p(n),true> $")+2pt}, text centered
              }]
  \node[main node] (1) {$\scriptsize\langle\texttt{p($n$)}, \true \rangle \break {\where}\, M$};
  \node[main node] (2) [below left=0.6cm and -1.5cm of 1] {$\scriptsize\langle\texttt{q($n$)}, \true \rangle \break {\where}\, M $};
  \node[main node] (3) [below right=0.6cm and -1.5cm of 1] {$\scriptsize\langle\texttt{r($n$)}, \true \rangle \break {\where}\, M$};

  \path
    (1) edge [shorten >=-1pt, shorten <=3pt] node [midway,left] {\scriptsize ${ r_1}$} (2)
    (1) edge [shorten >=-1pt, shorten <=3pt] node [midway,right] {\scriptsize$r_2$} (3);

\begin{scope}[xshift = 4.3 cm]
  \node[main node] (1) {$\scriptsize\langle\texttt{p($n$)}, \true \rangle \break \,{\where}\, M_1$};
  \node[main node] (2) [below left=0.6cm and -1.5cm of 1] {$\scriptsize\langle\texttt{q($n$)}, \true \rangle \break \,{\where}\, M_1$};
  \node[main node] (3) [below right=0.6cm and -1.5cm of 1] {$\scriptsize\langle\texttt{r($n$)}, \true \rangle \break \,{\where}\, M_1$};

  \path
    (1) edge [shorten >=-1pt, shorten <=3pt] node [midway,left] {\scriptsize ${ r_1}$} (2)
    (1) edge [shorten >=-1pt, shorten <=3pt] node [midway,right] {\scriptsize$r_2$} (3)
    (2) edge [dashed, shorten >=-2pt, shorten <=0pt] node [above] {\small$r_4$} (3);  

\end{scope}
\begin{scope}[xshift = 8.3 cm]
  \node[main node] (1) {$\scriptsize\langle\texttt{p($n$)}, \true \rangle \break \,{\where}\, M_2$};
  \node[main node] (2) [below left=0.6cm and -1.5cm of 1] {$\scriptsize\langle\texttt{q($n$)},\true \rangle \break \,{\where}\, M_2$};
  \node[main node] (3) [below right=0.6cm and -1.5cm of 1] {$\scriptsize\langle\texttt{r($n$)},\true \rangle \break \,{\where}\, M_2$};

  \path
    (1) edge [shorten >=-1pt, shorten <=3pt] node [midway,left] {\scriptsize ${ r_1}$} (2)
    (1) edge [shorten >=-1pt, shorten <=3pt] node [midway,right] {\scriptsize$r_2$} (3)
    (3) edge [dashed, shorten >=-2pt, shorten <=0pt] node [above] {\small$r_3$} (2);  

\end{scope}    
\end{tikzpicture}
\end{center}
%
%
According to Lemma~\ref{lm:confl-meta-level-join} shown below, the program is confluent.
\end{example}
When, furthermore, a state equivalence $\approxlogic$ is assumed,
we need also show joinability of $\beta$ corners, i.e., those composed by an equivalence and a transition.

\begin{definition}\label{def:critical-meta-level-corner:beta}
Let $\nonamerule H{H'}GC$ be a (lifted version of a) variant of a rule with  local variables $L$.
Assume
$S^+$, $B^+$ and $\Sigma_1$ are fresh meta-variables, and let
\begin{eqnarray*}
\Sigma_0 & \quad=\quad & \langle H{\uplus} H' {\uplus} S^+,B^+\rangle
\qquad\qquad\qquad
\Sigma_2  \quad=\quad  \langle H{\uplus} C {\uplus} S^+,G{\land}B^+\rangle\\
M  & \quad=\quad & \cinv(\Sigma_0)\land\willsucced B\land\willsucced(B{\rightarrow}\exists LG)\land
\text{\rm\texttt{freshVars($L$,$\Sigma_0$)}} \land \cequiv(\Sigma_0,\Sigma_1)
\end{eqnarray*}
When $(\Sigma_0\where M)$ is consistent, the following is \emph{a critical meta level $\beta$  corner
by rule application}.
$$(\Sigma_1\where M)\approxmetalogic(\Sigma_0\where M)\ourlongmapstometalogic(\Sigma_2\where M)$$
Let $b$ be a (lifted version of a) built-in atom
whose arguments are fresh variables.
Assume $S^+$, $B^+$ and $\Sigma_1$  are fresh meta-variables, and let
\begin{eqnarray*}
\Sigma_0 & \quad=\quad & \langle \{b\} {\uplus} S^+,B^+\rangle
\qquad\qquad\qquad
\Sigma_2  \quad=\quad  \langle S^+,b{\land}B^+\rangle\\
M  & \quad=\quad & \cinv(\Sigma_0)\land\willsucced B\land
\text{\rm\texttt{freshVars($L$,$\Sigma_0$)}} \land \cequiv(\Sigma_0,\Sigma_1)
\end{eqnarray*}
When $(\Sigma_0\where M)$ is consistent, the following is \emph{a critical meta level $\beta$  corner by built-in}.
$$(\Sigma_1\where M)\approxmetalogic(\Sigma_0\where M)\ourlongmapstometalogic(\Sigma_2\where M)$$
\end{definition}

\begin{lemma} \label{lm:confl-meta-level-join}
Let a terminating CHR program $\Pi$ with invariant $\Ilogic$ (and state quivalence $\approxlogic $) be given.
Then $\Pi$ is confluent (modulo $\approxlogic$) if and only if its set of critical corners (Def.s~\ref{def:critical-meta-level-corner:alpha}--\ref{def:critical-meta-level-corner:beta})
is split-joinable w.r.t.\ $\Imetalogic$ (modulo $\approxmetalogic$).
\end{lemma}

%
%
%
\begin{example}[Cont.\ Ex.~\ref{ex:set-critical-corners-logic}; adapted from~\cite{DBLP:journals/fac/ChristiansenK17}]
\label{ex:set-critical-corners-logicCont}
The invariant is formalized at the meta level as states of the form
{\small$$\langle \{\texttt{set($L$)}\}{\uplus}S,\true\rangle \where \texttt{type(constList,$L$)}{\land}\texttt{type(constItems,$S$)};$$}\noindent
we assume types \texttt{const} for all constants,
\texttt{constList} for all lists of such, and \texttt{constItems} for sets of constraints of the form \texttt{item($c$)}
where $c$ is a constant.

The state equivalence is formalized at the meta level as the
relationships of states of the following form, where \texttt{perm($L_1$,$L_2$)} means
that $L_1$ and $L_2$ are lists being permutations of each other; and $M^{\approx}$ stands for
{\small$\texttt{type(constList,$L_1$)}{\land}
$ $ 
\texttt{type(constList,$L_1$)}{\land}\texttt{perm($L_1$,$L_2$)}
{\land}\texttt{type(constItems,$S$)}$},
{\small$$\langle \{\texttt{set($L_1$)}\}{\uplus}S,\true\rangle \where M^{\approx}
\quad\approxmetalogic\quad
\langle \{\texttt{set($L_2$)}\}{\uplus}S,\true\rangle \where M^{\approx}
$$}\noindent
The critical object level corner 
with two set constraints in the states does not give rise to a critical meta level corner as the invariant is not satisfied.
The other one is shown here, including (with dotted arrows) its proof of joinability modulo equivalence; 
$M^\alpha$ stands for {\small $\texttt{type(const,$x_1$)}\land\texttt{type(constList,$L$)}\land\texttt{type(const,$x_2$)}
\land\break\texttt{type(constItems,$S$)}$}.
\begin{center}
{\tikzset{|/.tip={Bar[width=.8ex,round]}}

\noindent\begin{tikzpicture}[|->, auto,node distance=2cm and 1.0cm,
               main node/.style={font=\scriptsize             
              }]

  \node[main node] (1) {$\langle\{\texttt{item($x_1$)}, \texttt{set($L$)}, \texttt{item($x_2$)}\}{\uplus}S,\true\rangle\where M^\alpha$};
  \node[main node] (2) [below left=0.2cm and -3.1cm of 1] {$\langle\{\texttt{set([$x_1$|$L$])}, \texttt{item($x_2$)}\}{\uplus}S,\true\rangle\where M^\alpha$};
  \node[main node] (3) [below right=0.2cm and -3.1cm of 1] {$\langle\{\texttt{item($x_1$)},\texttt{set([$x_2$|$L$])},\}{\uplus}S,\true\rangle\where M^\alpha$};
  \node[main node] (2A) [below =0.4cm of 2] {$\langle\{\texttt{set([$x_2$,$x_1$|$L$])}\}{\uplus}S,\true\rangle\where M^\alpha$};
  \node[main node] (3A) [below =0.4cm of 3] {$\langle\{\texttt{set([$x_1$,$x_2$|$L$])}\}{\uplus}S,\true\rangle\where M^\alpha$};
  \path
    (1) edge [shorten >=-1pt, shorten >=3pt, shorten <=3pt]  (2)
    (2) edge [densely dashed, shorten >=-3pt, shorten <=-2pt]  (2A)
    (1) edge [shorten >=-1pt, shorten >= 3pt, shorten <=3pt]  (3) 
    (3) edge [densely dashed, shorten >=-3pt, shorten <=-2pt]  (3A);
\draw[-, dash pattern=on 2pt off 1pt, transform canvas={yshift=1pt},decorate, decoration={snake, segment length=7pt, amplitude=0.3mm}] (2A) -- (3A);
\draw[-, dash pattern=on 2pt off 1pt, transform canvas={yshift=-1pt},decorate, decoration={snake, segment length=7pt, amplitude=0.3mm}] (2A) -- (3A);
\end{tikzpicture}
}
\end{center}
%
%
We consider the following critical meta level $\beta$ corner.
$M^\beta$ stands for
{\small $\texttt{type(const,$x$)}
$\break$ 
{\land}\texttt{type(constList,$L_1$)}
{\land}\texttt{type(constList,$L_2$)}
{\land}
\texttt{perm($L_1$,$L_2$)}
{\land}\texttt{type(constItems,$S$)}$}.\\
\begin{center}
{\tikzset{|/.tip={Bar[width=.8ex,round]}}
\usetikzlibrary{decorations.pathmorphing}
\noindent\begin{tikzpicture}[|->, auto,node distance=2cm and 1.0cm,
               main node/.style={font=\scriptsize ,
               minimum width={5cm}, align=center        
              }]

  \node[main node] (1) {$\langle\{\texttt{item($x$)}, \texttt{set($L_1$)}\}{\uplus}S,\true\rangle\where M^\beta$};
  \node[main node] (2) [below left=0.2cm and -2.5cm of 1] {$\langle\{\texttt{item($x$)}, \texttt{set($L_2$)}\}{\uplus}S,\true\rangle\where M^\beta$};
  \node[main node] (3) [below right=0.2cm and -2.5cm of 1] {$\langle\{\texttt{set([$x$|$L_1$])}\}{\uplus}S,\true\rangle\where M^\beta$};
    \node[main node] (4) [below right=0.2cm and -2.5cm of 2] {$\langle\{\texttt{set([$x$|$L_2$])}\}{\uplus}S,\true\rangle\where M^\beta$};

\draw[-,  transform canvas={yshift=1pt,xshift=-0.4pt},decorate, decoration={snake, segment length=7pt, amplitude=0.3mm}] (1) -- (2); 
\draw[-,  transform canvas={yshift=-1pt},decorate, decoration={snake, segment length=7pt, amplitude=0.3mm}] (1) -- (2);     
   \path   
    (2) edge [densely dashed, shorten <=3pt]  (4)
    (1) edge [shorten >=-1pt, shorten <=3pt]  (3) ;

\draw[-, dash pattern=on 2pt off 1pt, transform canvas={yshift=1pt,xshift=-0.25pt},decorate, decoration={snake, segment length=7pt, amplitude=0.3mm}] (3) -- (4); 
\draw[-, dash pattern=on 2pt off 1pt, transform canvas={yshift=-1pt},decorate, decoration={snake, segment length=7pt, amplitude=0.3mm}] (3) -- (4); 
 
\end{tikzpicture}
}
%
%
\end{center}
\noindent All critical corners are joinable modulo equivalence, and since the program is obviously terminating,
Lemma~\ref{lm:confl-meta-level-join} gives that the program is confluent mod.\ equiv.
\end{example}
%
%
%
%
%
%
%
%

\section{Conclusion}\label{sec:conclusion}
\vskip -6pt
\noindent 
We generalized the critical pair approach using a meta level simulation to prove confluence under invariant and modulo equivalence for
Constraint Handling Rules. 
%
We have demonstrated how this principle makes it possible to express natural invariants and equivalences,
that cannot be expressed in CHR itself, in a formal way at the meta level, anticipating machine supported proofs using
a meta level constraint solver, based on a ground representation.
%
A constraint solver is currently under development,
partly inspired by~\cite{DBLP:journals/jlp/Christiansen98}.
Depending on the complexity of the invariants and equivalences -- and of the CHR programs under investigation --
it may be difficult to obtain a complete solver. 

For simplicity of notation, we did not include mechanisms to prevent  loops caused by propagation
rules;~\cite{DBLP:journals/fac/ChristiansenK17} has included this in a meta level representation for the Prolog based semantics,
and is easily adapted for the logic based semantics exposed in the present paper.

For comparison with
earlier work on confluence for CHR, we used here a logic-based CHR semantics, which has nice theoretical properties,
but is incompatible with standard implementations of CHR and applies only for a limited set of programs.
In~\cite{WFLP2018}, we have defined meta level constraints and a simulation for an alternative CHR semantics~\cite{DBLP:conf/lopstr/ChristiansenK14,DBLP:journals/fac/ChristiansenK17} that
reflects CHR's Prolog based implementation, including a correct handling of Prolog's non-logical devices
(e.g., \texttt{var}/1, \texttt{nonvar}/2, \texttt{is}/2) and runtime errors.
%
%

We could argue that the abstract simulations used for the classical CHR confluence results are special cases of abstract interpretations. When invariants are introduced -- or  when considering full CHR including Prolog-style non-logical devices, cf.~\cite{WFLP2018} -- this correspondence does not hold.

The concept of abstract simulations and their use for proving confluence (mod.\ equiv.) seem obvious to investigate for a large variety of rewrite based systems, e.g., constrained term rewriting, conditional term rewriting, interactive theorem provers, and rule-based specifications of abstract algorithms.

\subsection*{Acknowledgement}
We thank the anonymous reviewers for their insightful and detailed comments, 
suggesting to compare with a transformational approach, cf.\ Example~\ref{ex:transformationalapproach}, and helping us to clarify the relationship between abstract simulation and abstract interpretation.

\bibliography{CHR}

\end{document}